\documentclass[aps,prd,tightenlines,preprint,showpacs,floatfix,%
amssymb,byrevtex,nofootinbib]{revtex4}

\usepackage{amssymb}
\usepackage{graphicx,bm}

\begin{document}

\preprint{hep-ph/0702126}

%%%%%%%%%%%%%%%%%%%%% Title %%%%%%%%%%%%%%%%%%%%%%

\title{\boldmath $\Xi$ and $\Omega$ baryons in the Skyrme model}

%%%%%%%%%%%%%%%%%%%% Authors %%%%%%%%%%%%%%%%%%%%%
%%%%%%%%%%%%%%%%%%%% Addresses %%%%%%%%%%%%%%%%%%%%%

\author{Yongseok Oh}%
\email{yoh@comp.tamu.edu}

\affiliation{Department of Physics and Astronomy,
University of Georgia, Athens, GA 30602, U.S.A.}
\affiliation{
Cyclotron Institute and Physics Department, Texas A\&M
University, College Station, TX 77843, U.S.A.}

\date{\today}

%%%%%%%%%%%%%%%%%%%% Abstract %%%%%%%%%%%%%%%%%%%%%

\begin{abstract}

The mass spectrum and magnetic moments of $\Xi$ and $\Omega$ baryon
resonances are investigated in the bound state approach of the Skyrme model.
The empirical hyperon spectrum shows that several hyperon resonances
share a pattern of (approximately) equal mass spacings between the states
of same spin but of opposite parity.
It is found that this pattern can be explained mostly by the energy
difference between the $P$-wave and $S$-wave kaons bound to the soliton.
Although one cannot exclude the possibility that these states can be
described as pion-hyperon resonances, the present approach predicts that
$\Xi(1620)$ and $\Xi(1690)$ have $J^P = \frac12^-$, while $\Xi(1950)$ has
$J^P = \frac12^+$.
The differences with the quark model predictions for the $\Xi$ and $\Omega$
baryon spectrum are pointed out.
Several relations for the masses and magnetic moments of those
resonances are also obtained.

\end{abstract}


\maketitle

\section{Introduction}

In spite of the early efforts for studying $\Xi$ and $\Omega$ resonances,  
our understanding on those resonances is still far from complete.
Any new significant information on the multi-strangeness baryons has not
been accumulated during the last two decades \cite{PDG06} except for the
measurement of the magnetic moment of $\Omega^-(1672)$ \cite{WBCG95}.
(See, e.g., Ref.~\cite{WA89-99} for a latest experimental study on $\Xi$
resonances.)
Recently, however, interest in cascade physics is increasing.
New activities in measuring weak decays of $\Xi^0$ hyperon were
reported by the KTeV Collaboration \cite{KTeV05} and by the NA48/1
Collaboration \cite{NA48/1-07}.
Also the recent research activities on $\Xi$ resonances in relativistic
heavy-ion collisions can be found in Ref.~\cite{STAR-07}.
In addition, the cascade physics program of the CLAS Collaboration at
the Thomas Jefferson National Accelerator Facility has been launched and some
preliminary results were already reported
\cite{PDGN05,CLAS04d,CLAS06d,CLAS07b,cascade}.
The purpose of this program includes searching for and confirming
$\Xi$ resonances as well as understanding their properties
through electromagnetic production processes.
Investigation of the production mechanisms of $\Xi$ baryon production
induced by photon-nucleon reaction was just started
recently~\cite{cascade,LK03c,NOH06}.

At present, there are eleven $\Xi$ baryons and four $\Omega$ baryons
listed in the review of the Particle Data Group (PDG) \cite{PDG06}.
Among them only the octet and decuplet ground states, $\Xi(1318)$,
$\Xi(1530)$, and $\Omega(1672)$, have four-star ratings with definite
spin-parity,%
\footnote{Some quantum numbers of the ground states have not
been directly measured but assigned \cite{PDG06}.
The most recent measurement on the
spin of $\Omega^-(1672)$ was reported in Ref.~\cite{BABAR06}.}
and there are four $\Xi$ baryons and one $\Omega$ baryon
with three-star ratings.
Among the three-star-rated baryons, $\Xi(1820)$ is the only state whose
spin-parity quantum numbers are known and nothing is known about the
spin-parity of all the other states.
A summary on the $\Xi$ and $\Omega$ spectrum of PDG is given in
Table~\ref{tab:pdg}.

%%%%%%%%%%%%%%%%%% TABLE I
\begin{table}[t]
\centering
\begin{tabular}{ccc|ccc} \hline\hline
Particle & $I(J^P)$ & rating & Particle & $I(J^P)$ & rating \\
\hline\hline
%$\Sigma(1193)$ & $1(\frac12^+)$ & **** &
$\Xi(1318)$ & $\frac12(\frac12^+)$ & **** &
$\Omega(1672)$ & $0(\frac32^+)$ & ****
\\
%$\Sigma(1385)$ & $1(\frac32^+)$ & **** &
$\Xi(1530)$ & $\frac12(\frac32^+)$ & **** &
$\Omega(2250)$ & $0(?^?)$ & ***
\\
%$\Sigma(1480)$ & $1(?^?)$ & * &
$\Xi(1620)$ & $\frac12(?^?)$ & * &
$\Omega(2380)$ & $?(?^?)$ & **
\\
%$\Sigma(1560)$ & $1(?^?)$ & ** &
$\Xi(1690)$ & $\frac12(?^?)$ & *** &
$\Omega(2470)$ & $?(?^?)$ & **
\\
%$\Sigma(1580)$ & $1(\frac32^-)$ & ** &
$\Xi(1820)$ & $\frac12(\frac32^-)$ & *** &
\\
%$\Sigma(1620)$ & $1(\frac12^-)$ & ** &
$\Xi(1950)$ & $\frac12(?^?)$ & *** &
\\
%$\Sigma(1660)$ & $1(\frac12^+)$ & *** &
$\Xi(2030)$ & $\frac12(\ge \frac52^?)$ & *** &
\\
%$\Sigma(1670)$ & $1(\frac32^-)$ & **** &
$\Xi(2120)$ & $\frac12(?^?)$ & * &
\\
%$\Sigma(1690)$ & $1(?^?)$ & ** &
$\Xi(2250)$ & $\frac12(?^?)$ & ** &
\\
%$\Lambda(1116)$ & $0(\frac12^+)$ & **** &
$\Xi(2370)$ & $\frac12(?^?)$ & ** &
\\
%$\Lambda(1405)$ & $0(\frac12^-)$ & **** &
$\Xi(2500)$ & $\frac12(?^?)$ & * &
\\
\hline\hline
\end{tabular}
\caption{$\Xi$ and $\Omega$ baryons listed in the review of the
Particle Data Group~\cite{PDG06}.}
\label{tab:pdg}
\end{table}

Theoretically, there have been studies on the excited
multi-strangeness hyperons using phenomenological models. 
Those models could reproduce the octet and decuplet ground states,
but have very different and even contradictory predictions on the spectrum
of excited states.
Early studies in quark models are summarized in
Refs.~\cite{PBBF70,SGM74,Horgan74}.
A more detailed study including the quark dynamics was done by Chao,
Isgur, and Karl~\cite{CIK81} in a non-relativistic quark model.
In this model, the $\Xi(1820)\frac32^-$ is well explained and the third
lowest state following $\Xi(1318)$ and $\Xi(1530)$ is predicted to be at
a mass of 1695 MeV with $J^P = \frac12^+$.
Therefore, this state would be a good candidate for the observed
three-star resonance $\Xi(1690)$.
However, this expectation is not valid in other models that predict very
different $\Xi$ spectrum except for the ground states.
For example, in the relativized quark model of Capstick and Isgur
\cite{CI86}, the first excited state of $\Xi(\frac12^+)$ would have a
mass of around $1840$ MeV.
Furthermore, the third lowest $\Xi$ state would have a mass of $1755$ MeV
with $J^P = \frac12^-$, which is $65$ MeV larger than the mass of
$\Xi(1690)$.

In the one-boson exchange model of Glozman and Riska~\cite{GR96b}, the
third lowest state would have odd parity at a mass of 1758 MeV with
$J=1/2$ or $3/2$.
This mass is in the middle of the masses of $\Xi(1820)$ and $\Xi(1690)$.
Therefore, this model overestimates the mass of $\Xi(1690)$ and
underestimates that of $\Xi(1820)$.
(See also Refs.~\cite{GPVW97,VGV05}.)
As a result, the non-relativistic quark model \cite{CIK81}, relativized
quark model \cite{CI86}, and one-boson exchange model \cite{GR96b} give
very different predictions on the $\Xi$ spectrum.

The large $N_c$ QCD predictions, where $N_c$ is the number of colors, also
lead to a quite different $\Xi$ spectrum \cite{CC00,SGS02,GSS03,MS04b,MS06b}.
In this approach, the mass of $\Xi(1820)\frac32^-$ is used as an input and
the third lowest state would have $J^P = \frac12^-$ at a mass of 1780 MeV.
Although the lowest state with $J^P = \frac12^-$ of this model lies close
to the predictions on the same $J^P$ states of the quark
models \cite{CIK81,GR96b}, the explanation for the observed $\Xi(1690)$ is
still uncertain.
(A connection between the quark models and the large $N_c$ expansion was
recently discussed in Ref.~\cite{SBMS07}.)

In the algebraic model of Bijker {\it et al.\/} \cite{BIL00}, the third lowest
state would have $J^P = \frac12^+$ at a mass around $1730$ MeV.
It also predicts two $\Xi(\frac32^-)$ states which lie close to the
observed $\Xi(1820)$.
This model predicts richer hyperon spectrum than quark models and it has
very different predictions especially on the $J^P = \frac12^-$ states.

%%%%%%%%%%%%%%%%%% TABLE II
\begin{table*}[t]
\centering
\begin{tabular}{c|cccccc} \hline\hline
State & CIK \cite{CIK81} & CI \cite{CI86} & GR \cite{GR96b} &
Large-$N_c$ \cite{CC00,SGS02,GSS03,MS04b,MS06b} & BIL \cite{BIL00} &
QCD-SR \cite{LL02} (\cite{JO96}) \\ \hline
$\Xi(\frac12^+)$ & $1325$ & $1305$ & $1320$ &        & $1334$ &
$1320$ ($1320)$ \\
                 & $1695$ & $1840$ & $1798$ & $1825$ & $1727$ &    \\
                 & $1950$ & $2040$ & $1947$ & $1839$ & $1932$ &    \\ \hline
$\Xi(\frac32^+)$ & $1530$ & $1505$ & $1516$ &        & $1524$ &    \\
                 & $1930$ & $2045$ & $1886$ & $1854$ & $1878$ &    \\
                 & $1965$ & $2065$ & $1947$ & $1859$ & $1979$ &    \\ \hline
$\Xi(\frac12^-)$ & $1785$ & $1755$ & $1758$ & $1780$ & $1869$ &
$1550$ ($1630)$ \\
                 & $1890$ & $1810$ & $1849$ & $1922$ & $1932$ &    \\
                 & $1925$ & $1835$ & $1889$ & $1927$ & $2076$ &    \\ \hline
$\Xi(\frac32^-)$ & $1800$ & $1785$ & $1758$ & $1815$ & $1828$ &
$1840$ \\
                 & $1910$ & $1880$ & $1849$ & $1973$ & $1869$ &    \\
                 & $1970$ & $1895$ & $1889$ & $1980$ & $1932$ &    \\
\hline
$\Omega(\frac12^+)$ & $2190$ & $2220$ & $2068$ & $2408$ & $2085$ & \\
                    & $2210$ & $2255$ & $2166$ &        & $2219$ & \\ \hline
$\Omega(\frac32^+)$ & $1675$ & $1635$ & $1651$ &        & $1670$ & \\
                    & $2065$ & $2165$ & $2020$ & $1922$ & $1998$ & \\
                    & $2215$ & $2280$ & $2068$ & $2120$ & $2219$ & \\ \hline
$\Omega(\frac12^-)$ & $2020$ & $1950$ & $1991$ & $2061$ & $1989$ & \\ \hline
$\Omega(\frac32^-)$ & $2020$ & $2000$ & $1991$ & $2100$ & $1989$ & \\
\hline\hline
\end{tabular}
\caption{Low-lying $\Xi$ and $\Omega$ baryon spectrum of spin $1/2$ and
$3/2$ predicted by the non-relativistic quark model of
Chao {\it et al.\/} (CIK), relativized quark model of Capstick and Isgur
(CI), Glozman-Riska model (GR), large $N_c$ analysis,
algebraic model (BIL), and QCD sum rules (QCD-SR).
The mass is given in the unit of MeV.}
\label{tab:xi-omega}
\end{table*}

There also have been some efforts to construct QCD sum rules for $\Xi$
baryons, but the results depend strongly on the approximations and
assumptions made during the calculation \cite{JO96,LL02}.
The model of Ref.~\cite{JO96} prefers $\Xi(1690)$ as a $J^P
= \frac12^-$ state, while that of Ref.~\cite{LL02} predicts very low
mass for this state although it can describe $\Xi(1820)\frac32^-$.
The lattice QCD calculation for multi-strangeness baryons is still at
the very early stage.
But some progress has been reported, e.g., in Refs.~\cite{LDDH02,BGGH06}.

One common characteristic of the quark models is that it is very
difficult to accommodate $\Xi(1620)$ and $\Xi(1690)$ together.
In particular, the low mass of one-star-rated $\Xi(1620)$ is puzzling in
the quark models if its existence is confirmed by future experiments.
In Ref.~\cite{AASW03}, Azimov {\it et al.\/} assigned $J^P = \frac12^-$
to $\Xi(1620)$ so that it can form a narrow light baryon octet with
$N'(\sim\!\! 1100)$, $\Lambda(1330)$, and $\Sigma(1480)$.
Although this model satisfies the Gell-Mann--Okubo mass relation, it
requires the existence of very low mass nucleon and $\Lambda$
resonances.
In Ref.~\cite{ROB02}, Ramos {\it et al.\/}, suggested to identify
$\Xi(1620)$ as a dynamically generated $S$-wave $\Xi$ resonance based on
a unitary extension of chiral perturbation theory, which predicts a
$\Xi$ resonance at a mass around $1606$ MeV.
Then $\Xi(1620)$ has $J^P = \frac12^-$ and it was suggested to form a
unitary octet with $N(1535)$, $\Lambda(1670)$, and $\Sigma(1620)$,
which, however, does not satisfy the Gell-Mann--Okubo mass formula.
In a similar approach but with different details,
Garc{\'\i}a-Recio {\it et al.\/} \cite{GLN04} claimed that both $\Xi(1620)$
and $\Xi(1690)$ are $S$-wave resonances having $J^P = \frac12^-$, while
$\Xi(1690)$ was claimed to be a genuine resonance in Ref.~\cite{ROB02}.
(Some $\Omega$ baryon resonances are discussed in a similar approach in
Ref.~\cite{GNS06}.)

The predictions of various models on the $\Xi$ spectrum are summarized in
Table~\ref{tab:xi-omega}, which evidently shows strong model dependence of the
resulting spectrum. (See also Refs.~\cite{TZ86,BBHM90,CG99} for the other model
predictions.)
The strong model dependence can also be found in the $\Omega$ baryon
spectrum as presented in the same Table, which shows the degeneracy of $J^P =
\frac12^-$ and $J^P=\frac32^-$ $\Omega$ baryons in the quark models of
Refs.~\cite{CIK81,GR96b,BIL00}.

In this paper, we investigate strange hyperon resonances in the bound state
approach of the Skyrme model \cite{CK85}.
In this approach, strange hyperons are described as the bound states of
the soliton and $K/K^*$ meson(s).
This model was successful to describe the ground state
octet and decuplet hyperons \cite{CHK88,SMNR89}.
It also has been applied to heavy-quark baryons
\cite{RRS90,RRS92,OMRS91} and the resulting heavy baryon spectrum was
shown to respect the heavy quark spin symmetry once the heavy
vector meson fields were treated in a consistent way
\cite{JMW93a,GMSS93,OPM94c,MOPR95,OP96b}.
Recent discussion on this model for the existence of exotic states can
be found in Refs.~\cite{IKOR03,PRM04}.
In the beginning, this model was developed with Lagrangians which have
pseudoscalar $K$ meson only by integrating out the $K^*$ vector meson.
However, as discussed in Ref.~\cite{PRM04}, although the $K^*$ vector
meson has a larger mass, it may have nontrivial role, which cannot be
correctly accounted for if it is integrated out in favor of pseudoscalar
meson.
But our discussion in this paper does not depend on the choice of the
degrees of freedom in the strangeness direction and from now on by
``kaons'' we mean both $K$ and $K^*$ mesons.

One novel feature of the bound state model is that it renders two kinds
of bound kaons: one in $P$-wave and one in $S$-wave \cite{CK85}.%
\footnote{For soliton--heavy-meson systems, there are more bound states
than the soliton-kaon system \cite{OP95}.}
Therefore, whereas simple quark models have difficulties in describing
$\Lambda(1405)$, this model gives a natural description of $\Lambda(1116)$ and
$\Lambda(1405)$ on the same ground \cite{SSG95}.
Namely, the even-parity $\Lambda(1116)$ is a bound
state of the soliton and $P$-wave kaon, while the odd-parity
$\Lambda(1405)\frac12^-$ contains an $S$-wave kaon.
The mass difference between these two $\Lambda$ hyperons is $289$ MeV.
One can easily find similar mass gaps in hyperon mass spectrum, e.g., between
$\Sigma(1385)\frac32^+$ and $\Sigma(1670)\frac32^-$ and between
$\Xi(1530)\frac32^+$ and $\Xi(1820)\frac32^-$.
Their mass differences are $285$ MeV and $290$ MeV, respectively, and
are very close to the mass difference between $\Lambda(1405)$
and $\Lambda(1116)$.
This observation gives us a strong motivation for the application of this
model to low-lying hyperon resonances.
The application to excited multi-strangeness baryons was first
tried in Refs.~\cite{DNR89,RRS92} by taking the simplest Skyrme model
Lagrangian for the dynamics of the soliton-kaon system.
However, the results were not so successful mainly because of the
simplicity of the adopted Lagrangian.
In this paper, we do not work with a specific Lagrangian for the system
of the soliton and kaon.
Instead, we focus on the general mass formula of this approach and 
find mass and magnetic moment sum rules among the even and odd parity
baryons.
For this purpose, we develop an improved mass formula which
differs from that given in, e.g., Ref.~\cite{SMNR89}.
We will also fit the mass parameters to some known baryon masses and make
predictions on the hyperon spectrum.
It should be mentioned, however, that the present approach cannot
be applied to \emph{all} hyperon resonances.
There exist many low-lying hyperon resonances which potentially could be
described as resonances in the pion-hyperon channel.
In the Skyrme model, these resonances then may be investigated by including
fluctuating pion fields \cite{SWHH89}.
In this paper we focus on the multi-kaon--soliton channel for describing
several hyperon resonances leaving the issue on the pion-hyperon
resonances to a future study.

This paper is organized as follows.
In the next Section, we briefly review the mass formula of the
bound-state approach and develop a new formula.
In Sec.~III, we present our results for the hyperon mass spectrum.
Mass sum rules and the ``best fit'' results for the hyperon mass
spectrum will be given.
Section IV is devoted to the magnetic moments of baryons and we derive
several magnetic moment sum rules.
We conclude in Sec.~V.

\section{Quantization in bound-state model}

Referring the details to Refs.~\cite{MOPR95,OP96a}, we begin with a
brief review on the derivation of the mass formula of the bound-state
model.
We first consider the case of only one bound kaon.
Then the mass of a hyperon with isospin $i$ and spin $j$ in the soliton-kaon
bound state model reads
\begin{equation}
M(i,j) = M_{\rm sol} + \omega + E_{\rm rot},
\end{equation}
where $M_{\rm sol}$ is the soliton mass, $\omega$ the energy of the
bound kaon, and $E_{\rm rot}$ the rotational energy arising from the
collective rotation.
The rotational energy is obtained as
\begin{equation}
E_{\rm rot} = \frac{1}{2\mathcal{I}} \left( \bm{R} - \bm{\Theta}
\right)^2,
\end{equation}
where $\mathcal{I}$ is the moment of inertia of the soliton and
$\bm{\Theta}$ is the isospin of the kaon embedded in the
soliton field.
Effectively, $\bm{R}$ has the role of the spin of the light $u$, $d$ quarks.
This mass formula respects $1/N_c$ expansion since
$M_{\rm sol}$, $\omega$, and $E_{\rm rot}$ are of the order of $N_c$,
$N_c^0$, and $1/N_c$, respectively.
The collective quantization gives
\begin{eqnarray}
I^a &=& D^{ab} R^b, \nonumber \\
\bm{J} &=& \bm{R} + \bm{J}_K^{},
\end{eqnarray}
where $D^{ab}$ is the SU(2) adjoint representation associated with
the collective variables and
$\bm{J}_K^{}$ is the ``grand-spin'' of the kaon: $\bm{J}_K^{} = \bm{L}_K +
\bm{S}_K^{} + \bm{T}_K^{}$, with $\bm{L}_K$, $\bm{S}_K^{}$, and $\bm{T}_K^{}$
being the orbital angular momentum, spin, and isospin of the kaon.
It is well-known that the bound state of kaon is obtained only for
$j_K^{} = 1/2$~\cite{CK85}.
In order to represent $\bm{\Theta}$ in terms of good quantum
numbers, one makes use of the relation,
\begin{equation}
\bm{\Theta} = -c \bm{J}_K^{}.
\end{equation}
This defines the constant $c$ which plays the role of the hyperfine
splitting constant.
Therefore, as pointed out by Ref.~\cite{KK90}, the rotational energy is
very similar to the magnetic moment interactions of quark models since
$E_{\rm rot}$ contains the spin-spin interactions.
In this sense, the constant $c$ distinguishes the strength of the
interactions with strange quarks from that of the light-quark--light-quark
interactions.
When there are $n$ bound kaons of the same kind, the mass formula
of Ref.~\cite{SMNR89} gives
\begin{eqnarray}
M(i,j) &=& M_{\rm sol} + n\omega
+ \frac{1}{2\mathcal{I}} \left\{ c j(j+1) + (1-c)i(i+1)
\right. \nonumber \\ && \mbox{} \left.
+ c(c-1) j_1^{}(j_1^{}+1) \right\},
\label{eq:smnr}
\end{eqnarray}
where $j_1^{} = n j_K^{}$ being constrained by the Bose statistics.

If there are two kinds of bound mesons, such as the system including
both the $P$-wave and $S$-wave kaons or the system of $K$ meson and $D$
meson, the rotational energy reads
\begin{equation}
E_{\rm rot} = \frac{1}{2\mathcal{I}} \left( \bm{R} - \bm{\Theta}_1 -
\bm{\Theta}_2 \right)^2.
\end{equation}
By defining $\bm{J}_m = \bm{J}_1 + \bm{J}_2$, the mass formula
of Ref.~\cite{RRS90} gives

%\begin{widetext}

\begin{eqnarray}
M(i,j,j_m^{}) &=& M_{\rm sol} + n_1^{} \omega_1^{} + n_2^{} \omega_2^{}
\nonumber \\ && \mbox{}
+ \frac{1}{2\mathcal{I}} \Biggl\{
i(i+1) + c_1^{} c_2^{} j_m^{}(j_m^{}+1)
+ (c_1^{}-c_2^{}) [ c_1^{} j_1^{}(j_1^{}+1) - c_2^{} j_2^{} (j_2^{}+1) ]
\nonumber \\ && \mbox{} +
[j(j+1) - j_m^{}(j_m^{}+1) - i(i+1)]
\nonumber \\ && \mbox{} \times
\left[ \frac{c_1^{}+c_2^{}}{2}
+ \frac{c_1^{}-c_2^{}}{2} \frac{j_1^{}(j_1^{}+1) -
j_2^{}(j_2^{}+1)}{j_m^{}(j_m^{}+1)} \right]
\Biggr\},
\label{eq:mass1}
\end{eqnarray}
where $\omega_i^{}$, $j_i^{}$, $c_i^{}$ are the energy, grand-spin, and
hyperfine constant of the $i$-th kind of bound meson whose number
is represented by $n_i^{}$.%
\footnote{The typographical errors committed in Refs.~\cite{RRS90,RRS92}
were corrected in Ref.~\cite{OMRS91}.}

%\end{widetext}

However, the assumptions made to arrive at the mass formula
(\ref{eq:mass1}) require further consideration.
Firstly, it is assumed that
\begin{equation}
\bm{\Theta}^2 = c^2 \bm{J}_K^2.
\label{eq:theta2}
\end{equation}
As mentioned before, $\bm{R}$ is the isospin of the kaon embedded in the
soliton field.
It was pointed out in Ref.~\cite{OPM94c} that $\bm{\Theta}^2$ can be
exactly calculated in the infinite heavy mass limit thanks to the
simplicity of the heavy meson (spatial) wavefunction.
In this limit, $\bm{\Theta}$ is nothing but the isospin operator sandwiched by
$\bm{\tau} \cdot \hat{\bm{r}}$.
As a result, $\bm{\Theta}^2$ can be calculated exactly as
\begin{equation}
\bm{\Theta}^2 = \bm{J}_K^2 = \frac34,
\end{equation}
which evidently shows the failure of the approximation (\ref{eq:theta2})
in this limit.
It should also be mentioned that the role of vector mesons is crucial to
get the correct heavy quark limit.
With this observation, one can show that the soliton-meson bound state model
is equivalent to the nucleon-meson bound state model of
Ref.~\cite{JMW93a} in the heavy quark limit.
Another ambiguity in calculating $\bm{\Theta}^2$ is that the kaon-kaon
interactions which have been neglected so far should contribute to this
term.
Including the kaon-kaon interactions would have important role especially
for multi-strangeness baryons and $\bm{\Theta}^2$ may depend on the
strangeness of the baryon.
However, this requires to expand the soliton Lagrangian up to the kaon
quartic terms, which is laborious \cite{WK94} and beyond the scope of
this work.%
\footnote{The anharmonic corrections in the bound-state soliton model
were first addressed by Bj{\"o}rnberg {\it et al.\/} in Ref.~\cite{BDR95}
within a concise form of the effective Lagrangian.
By treating the kaon quartic terms as a perturbation, the authors found
that the anharmonic corrections would lead to an about 10\% correction
to the $P$-wave kaon energy.}
Instead of working with a specific model Lagrangian, we introduce
another parameter $\bar{c}$,
\begin{equation}
\bm{\Theta}^2 = \bar{c} \bm{J}_K^2,
\label{eq:theta2-new}
\end{equation}
as suggested by the authors of Refs.~\cite{KK90,WK94}.
Therefore the approximation (\ref{eq:theta2}) corresponds to $\bar{c} =
c^2$, which is shown to be comparable to the quark model with magnetic
moment interactions \cite{KK90,WK94}.
However, it should be kept in mind that, in the heavy quark limit, one has
$\bar{c} \to 1$ and $c \to 0$, which shows that the relation $\bar{c}
= c^2$ does not hold in general.
By working with Eq.~(\ref{eq:theta2-new}), the mass formula is still
in the form of Eq.~(\ref{eq:smnr}) but $c^2$ is replaced by $\bar{c}$.

Secondly, in order to arrive at the expression for the $(c_1^{} - c_2^{})$
term of Eq.~(\ref{eq:mass1}), $\bm{J}_1 - \bm{J}_2$ is assumed to precess about
the $\bm{J}_m$ direction, which is similar to the case of anomalous Zeeman
effect in weak magnetic field.
This assumption allows us to write
\begin{equation}
(\bm{J}_1 - \bm{J}_2) \cdot \bm{R} \approx
\frac{(\bm{J}_1 - \bm{J}_2) \cdot
\bm{J}_m}{\bm{J}_m^2} \bm{R} \cdot \bm{J}_m,
\label{eq:zeeman}
\end{equation}
which then leads to the geometrical factor of the $(c_1^{} - c_2^{})$
term.
However, this can be a good approximation only when the
vector $\bm{J}_1 - \bm{J}_2$ precesses about the $\bm{J}_m$ axis.
But, in fact, this term can be calculated directly from the wavefunctions
without making any assumptions.
In principle, this term can contribute only to the $\Xi$ and $\Omega$ baryon
masses but the expression of Eq.~(\ref{eq:zeeman}) vanishes in both
cases since
$j_1^{} = j_2^{}$ for $\Xi$ baryons and $r = 0$ for $\Omega$ baryons.
However, although this term vanishes for those baryons, it
causes a mixing between odd-parity $\Xi$ resonances as
will be shown in the next Section.
%\begin{widetext}
Therefore, we work with the mass formula,
\begin{eqnarray}
M(i,j,j_m^{}) &=& M_{\rm sol} + n_1^{} \omega_1^{} + n_2^{} \omega_2^{}
\nonumber \\ && \mbox{}
+ \frac{1}{2\mathcal{I}} \Biggl\{
i(i+1) + c_1^{} c_2^{} j_m^{}(j_m^{}+1)
+ (\bar{c}_1^{} - c_1^{} c_2^{}) j_1^{}(j_1^{}+1)
+ (\bar{c}_2^{} - c_1^{} c_2^{}) j_2^{}(j_2^{}+1)
\nonumber \\ && \mbox{} \qquad +
\frac{c_1^{}+c_2^{}}{2} [j(j+1) - j_m^{}(j_m^{}+1) - i(i+1)]
+ \frac{c_1^{} - c_2^{}}{2} \bm{R} \cdot (\bm{J}_1 - \bm{J}_2)
\Biggr\}.
\label{eq:mass0}
\end{eqnarray}
When there is only one kind of bound kaon, this formula reduces to
Eq.~(\ref{eq:smnr}) by replacing $c^2$ by $\bar{c}$.

%\end{widetext}

\section{Mass spectrum}

In order to calculate baryon mass spectrum, we need to construct the
baryon wavefunctions.
We classify the states according to the quantum numbers, $i$,
$j^P$, $j_1$, $j_2$, and $j_m$ as shown in Table~\ref{tab:qn}.
Since we are considering the $P$- and $S$-wave bound kaons, we can construct
two $\Lambda$'s, four $\Sigma$'s, seven $\Xi$'s and six $\Omega$'s which
lie below or close to the threshold energies.
Table~\ref{tab:qn} also gives the names of each state of which subscript
shows the spin-parity quantum numbers explicitly.
The number following the spin-parity in the subscript is introduced to
distinguish the states of the same spin-parity.
In particular, the subscript $0$ means that the baryon belongs to the octet
or decuplet ground state of the quark model.

%%%%%%%%%%%%%%%%%% TABLE III
\begin{table*}[t]
\centering
\begin{tabular}{cccccccc} \hline\hline
Strangeness & $n_{\ell=1}^{}$ & $n_{\ell=0}^{}$ & $j_m^{}$ & $i$ & $j^P$ &
Particle Name & Particle Data Group
\\ \hline
$S=0$ & $0$ & $0$ & $0$ & $1/2$ & $1/2^+$ & $N$ & $N(939)$ \\
      & $0$ & $0$ & $0$ & $3/2$ & $3/2^+$ & $\Delta$ & $\Delta(1232)$ \\
\hline
$S=-1$ & $1$ & $0$ & $1/2$ & $0$ & $1/2^+$ & $\Lambda_{1/2^+,0}$ &
$\Lambda(1116)$ \\
       & $0$ & $1$ & $1/2$ & $0$ & $1/2^-$ & $\Lambda_{1/2^-,1}$ &
$\Lambda(1405)$ \\
       & $1$ & $0$ & $1/2$ & $1$ & $1/2^+$ & $\Sigma_{1/2^+,0}$ &
$\Sigma(1193)$ \\
       & $1$ & $0$ & $1/2$ & $1$ & $1/2^+$ & $\Sigma_{3/2^+,0}$ &
$\Sigma(1385)$ \\
       & $0$ & $1$ & $1/2$ & $1$ & $1/2^-$ & $\Sigma_{1/2^-,1}$ &
\\
       & $0$ & $1$ & $1/2$ & $1$ & $3/2^-$ & $\Sigma_{3/2^-,1}$ &
\\ \hline
$S=-2$ & $2$ & $0$ & $1$ & $1/2$ & $1/2^+$ & $\Xi_{1/2^+,0}$ &
$\Xi(1318)$ \\
       & $2$ & $0$ & $1$ & $1/2$ & $3/2^+$ & $\Xi_{3/2^+,0}$ &
$\Xi(1530)$ \\
       & $1$ & $1$ & $0$ & $1/2$ & $1/2^-$ & $\Xi_{1/2^-,1}$ & \\
       & $1$ & $1$ & $1$ & $1/2$ & $1/2^-$ & $\Xi_{1/2^-,2}$ & \\
       & $1$ & $1$ & $0$ & $1/2$ & $3/2^-$ & $\Xi_{3/2^-,1}$ &
$\Xi(1820)$ \\
       & $0$ & $2$ & $1$ & $1/2$ & $1/2^+$ & $\Xi_{1/2^+,1}$ & \\
       & $0$ & $2$ & $1$ & $1/2$ & $3/2^+$ & $\Xi_{3/2^+,1}$ & \\
\hline
$S=-3$ & $3$ & $0$ & $3/2$ & $0$ & $3/2^+$ & $\Omega_{3/2^+,0}$ &
$\Omega(1672)$ \\
       & $2$ & $1$ & $1/2$ & $0$ & $1/2^-$ & $\Omega_{1/2^-,1}$ & \\
       & $2$ & $1$ & $3/2$ & $0$ & $3/2^-$ & $\Omega_{3/2^-,1}$ & \\
       & $1$ & $2$ & $1/2$ & $0$ & $1/2^+$ & $\Omega_{1/2^+,1}$ & \\
       & $1$ & $2$ & $3/2$ & $0$ & $3/2^+$ & $\Omega_{3/2^+,1}$ & \\
       & $0$ & $3$ & $3/2$ & $0$ & $3/2^-$ & $\Omega_{3/2^-,2}$ & 
\\ \hline\hline
\end{tabular}
\caption{Particles with quantum numbers. The last column shows the
particle identification with the octet and decuplet ground states as well
as $\Lambda(1405)$ and $\Xi(1820)$.}
\label{tab:qn}
\end{table*}

Given in Table~\ref{tab:states} are the spin-up wavefunctions of each
baryon in the basis of $|r,r_z^{} \rangle |j_1^{},
j_{1z}^{} \rangle_{\ell=1} |j_2^{}, j_{2z}^{} \rangle_{\ell=0}$.
It is then straightforward to write down the baryon masses and the
results are given in Table~\ref{tab:masses}.

%%%%%%%%%%%%%%%%%% TABLE IV
\begin{table*}[t]
\centering
\begin{tabular}{cc} \hline\hline
Particle & State \\ \hline
$|N \rangle$ & $|\frac12 \frac12 \rangle_R$
\\
$|\Delta \rangle$ & $|\frac32 \frac32 \rangle_R$
\\
$|\Lambda_{1/2^+,0} \rangle$ & $|0 0 \rangle_R | \frac12 \frac12
\rangle_1$
\\
$|\Sigma_{1/2^+,0} \rangle$ &
$- \frac{1}{\sqrt3} |10 \rangle_R |\frac12 \frac12 \rangle_1
+ \sqrt{\frac23} |1 1 \rangle_R | \frac12 - \frac12 \rangle_1$
\\
$|\Sigma_{3/2^+,0} \rangle$ & $| 1 1 \rangle_R | \frac12 \frac12
\rangle_1$
\\
$|\Lambda_{1/2^-,1} \rangle$ & $|0 0 \rangle_R | \frac12 \frac12
\rangle_0$
\\
$|\Sigma_{1/2^-,1} \rangle$ &
$- \frac{1}{\sqrt3} |10 \rangle_R |\frac12 \frac12 \rangle_0
+ \sqrt{\frac23} |1 1 \rangle_R | \frac12 - \frac12 \rangle_0$
\\
$|\Sigma_{3/2^-,1} \rangle$ & $| 1 1 \rangle_R | \frac12 \frac12
\rangle_0$
\\
$| \Xi_{1/2^+,0} \rangle$ & $\frac{1}{\sqrt3} | \frac12 \frac12
\rangle_R | 1 0 \rangle_1 - \sqrt{\frac23} | \frac12 -\frac12 \rangle_R
| 1 1 \rangle_1$
\\
$| \Xi_{3/2^+,0} \rangle$ & $| \frac12 \frac12 \rangle_R | 1 1
\rangle_1$
\\
$| \Xi_{1/2^-,1} \rangle$ & $\frac{1}{\sqrt2} | \frac12 \frac12
\rangle_R \left[ | \frac12 \frac12 \rangle_1 | \frac12 -\frac12
\rangle_0 - | \frac12 -\frac12 \rangle_1 | \frac12 \frac12
\rangle_0  \right]$
\\
$| \Xi_{1/2^-,2} \rangle$ & $\frac{1}{\sqrt6} | \frac12 \frac12
\rangle_R \left[ | \frac12 \frac12 \rangle_1 | \frac12 -\frac12
\rangle_0 + | \frac12 -\frac12 \rangle_1 | \frac12 \frac12
\rangle_0  \right]
- \sqrt{\frac23} | \frac12 - \frac12 \rangle_R | \frac12 \frac12
  \rangle_1  \frac12 \frac12 \rangle_0$
\\
$| \Xi_{3/2^-,1} \rangle$ & $| \frac12 \frac12 \rangle_R | \frac12
\frac12 \rangle_1 | \frac12 \frac12 \rangle_0$
\\
$| \Xi_{1/2^+,1} \rangle$ & $\frac{1}{\sqrt3} | \frac12 \frac12
\rangle_R | 1 0 \rangle_0 - \sqrt{\frac23} | \frac12 -\frac12 \rangle_R
| 1 1 \rangle_0$
\\
$| \Xi_{3/2^+,1} \rangle$ & $| \frac12 \frac12 \rangle_R | 1 1
\rangle_0$
\\
$| \Omega_{3/2^+,0} \rangle$ & $| 0 0 \rangle_R | \frac32 \frac32
\rangle_1$
\\
$| \Omega_{1/2^-,1} \rangle$ & $\frac{1}{\sqrt3}
| 0 0 \rangle_R \left[ \sqrt2 | 1 1 \rangle_1 | \frac12 - \frac12
\rangle_0 - | 1 0 \rangle_1 | \frac12 \frac12 \rangle_0 \right]$
\\
$| \Omega_{3/2^-,1} \rangle$ & $| 0 0 \rangle_R | 1 1 \rangle_1 |
\frac12 \frac12 \rangle_0$
\\
$| \Omega_{1/2^+,1} \rangle$ & $\frac{1}{\sqrt3}
| 0 0 \rangle_R \left[ -\sqrt2 | \frac12 - \frac12 \rangle_1 |1 1
\rangle_0 + | \frac12 \frac12 \rangle_1 | 1 0 \rangle_0 \right]$
\\
$| \Omega_{3/2^+,1} \rangle$ & $|0 0 \rangle_R | \frac12 \frac12
\rangle_1 | 1 1 \rangle_0$
\\
$| \Omega_{3/2^-,2} \rangle$ & $|0 0 \rangle_R | \frac32 \frac32
\rangle_0$
\\ \hline \hline
\end{tabular}
\caption{States of spin-up baryons in the bases of 
$|r,r_z^{} \rangle |j_1^{},
j_{1z}^{} \rangle_{\ell=1} |j_2^{}, j_{2z}^{} \rangle_{\ell=0}$.}
\label{tab:states}
\end{table*}

%%%%%%%%%%%%%%%%%% TABLE V
\begin{table*}[t]
\centering
\begin{tabular}{cc} \hline\hline
Particle & Mass \\ \hline
$N$ & $M_{\rm sol} + \frac{3}{8\mathcal{I}}$
\\
$\Delta$ & $M_{\rm sol} + \frac{15}{8\mathcal{I}}$
\\
$\Lambda_{1/2^+,0}$ & $M_{\rm sol} + \omega_1^{}
+ \frac{1}{2\mathcal{I}} \frac34 \bar{c}_1^{}$
\\
$\Lambda_{1/2^-,1}$ & $M_{\rm sol} + \omega_2^{}
+ \frac{1}{2\mathcal{I}} \frac34 \bar{c}_2^{}$
\\
$\Sigma_{1/2^+,0}$ & $M_{\rm sol} + \omega_1^{}
+ \frac{1}{2\mathcal{I}} \left( 2 + \frac34 \bar{c}_1^{} - 2c_1^{} \right)$
\\
$\Sigma_{3/2^+,0}$ & $M_{\rm sol} + \omega_1^{}
+ \frac{1}{2\mathcal{I}} \left( 2 + \frac34 \bar{c}_1^{} + c_1^{} \right)$
\\
$\Sigma_{1/2^-,1}$ & $M_{\rm sol} + \omega_2^{}
+ \frac{1}{2\mathcal{I}} \left( 2 + \frac34 \bar{c}_2^{} - 2c_2^{} \right)$
\\
$\Sigma_{3/2^-,1}$ & $M_{\rm sol} + \omega_2^{}
+ \frac{1}{2\mathcal{I}} \left( 2 + \frac34 \bar{c}_2^{} + c_2^{} \right)$
\\
$\Xi_{1/2^+,0}$ & $M_{\rm sol} + 2\omega_1^{}
+ \frac{1}{2\mathcal{I}} \left( \frac34 + 2 \bar{c}_1^{} - 2c_1^{} \right)$
\\
$\Xi_{3/2^+,0}$ & $M_{\rm sol} + 2\omega_1^{}
+ \frac{1}{2\mathcal{I}} \left( \frac34 + 2 \bar{c}_1^{} + c_1^{} \right)$
\\
$\Xi_{1/2^-,1}$ & $M_{\rm sol} + \omega_1^{} + \omega_2^{}
+ \frac{1}{2\mathcal{I}} \left( \frac34 + \frac34 \bar{c}_1^{}
- \frac32 c_1^{} c_2^{} + \frac34 \bar{c}_2^{} \right)$
\\
$\Xi_{1/2^-,2}$ & $M_{\rm sol} + \omega_1^{} + \omega_2^{}
+ \frac{1}{2\mathcal{I}} \left( \frac34 + \frac34 \bar{c}_1^{}
+ \frac12 c_1^{} c_2^{} + \frac34 \bar{c}_2^{} - c_1^{} - c_2^{} \right)$
\\
$\Xi_{3/2^-,1}$ & $M_{\rm sol} + \omega_1^{} + \omega_2^{}
+ \frac{1}{2\mathcal{I}} \left( \frac34 + \frac34 \bar{c}_1^{}
+ \frac12 c_1^{} c_2^{} + \frac34 \bar{c}_2^{} + \frac12 c_1^{}
+ \frac12 c_2^{} \right)$
\\
$\Xi_{1/2^+,1}$ & $M_{\rm sol} + 2\omega_2^{}
+ \frac{1}{2\mathcal{I}} \left( \frac34 + 2 \bar{c}_2^{} - 2c_2^{} \right)$
\\
$\Xi_{3/2^+,1}$ & $M_{\rm sol} + 2\omega_2^{}
+ \frac{1}{2\mathcal{I}} \left( \frac34 + 2 \bar{c}_2^{} + c_2^{} \right)$
\\
$\Omega_{3/2^+,0}$ & $M_{\rm sol} + 3\omega_1^{}
+ \frac{1}{2\mathcal{I}} \frac{15}{4} \bar{c}_1^{}$
\\
$\Omega_{1/2^-,1}$ & $M_{\rm sol} + 2\omega_1^{} + \omega_2^{}
+ \frac{1}{2\mathcal{I}} \left( 2 \bar{c}_1^{} - 2 c_1^{} c_2^{}
+ \frac34 \bar{c}_2^{} \right)$
\\
$\Omega_{3/2^-,1}$ & $M_{\rm sol} + 2\omega_1^{} + \omega_2^{}
+ \frac{1}{2\mathcal{I}} \left( 2 \bar{c}_1^{} + c_1^{} c_2^{}
+ \frac34 \bar{c}_2^{} \right)$
\\
$\Omega_{1/2^+,1}$ & $M_{\rm sol} + \omega_1^{} + 2\omega_2^{}
+ \frac{1}{2\mathcal{I}} \left( \frac34 \bar{c}_1^{} - 2 c_1^{} c_2^{}
+ 2 \bar{c}_2^{} \right)$
\\
$\Omega_{3/2^+,1}$ & $M_{\rm sol} + \omega_1^{} + 2\omega_2^{}
+ \frac{1}{2\mathcal{I}} \left( \frac34 \bar{c}_1^{} + c_1^{} c_2^{}
+ 2 \bar{c}_2^{} \right)$
\\
$\Omega_{3/2^-,2}$ & $M_{\rm sol} + 3\omega_2^{}
+ \frac{1}{2\mathcal{I}} \frac{15}{4} \bar{c}_2^{}$
\\ \hline\hline
\end{tabular}
\caption{Baryon masses in terms of mass parameters.
The quantities with the subscript $1$ are for the $P$-wave kaon and those
with the subscript $2$ are for the $S$-wave kaon.}
\label{tab:masses}
\end{table*}

For a given model Lagrangian, the mass parameters, namely, $M_{\rm sol}$,
$\mathcal{I}$, $\omega_{1,2}^{}$, $c_{1,2}^{}$, and $\bar{c}_{1,2}^{}$, can be
computed, and, therefore, these quantities are important to understand
the dynamics of the soliton-meson system.
However, our purpose is to test the mass formula without resorting to a
specific model Lagrangian.
We believe that by employing more realistic effective action for the
soliton-meson system one can calculate the mass parameters which converge to
the values of our estimation.
For this purpose, therefore, we fit the mass parameters to the known
baryon masses.
Firstly, the soliton mass and its moment of inertia can be obtained from
the nucleon and $\Delta$ masses as
\begin{equation}
M_{\rm sol} = 866 \mbox{ MeV}, \qquad
\mathcal{I} = 1.01 \mbox{ fm}.
\end{equation}
Secondly, the parameters for the $P$-wave kaon are obtained from
$\Lambda(1116)$, $\Sigma(1385)$, and $\Xi(1318)$, which give
\begin{equation}
\omega_1^{} = 211 \mbox{ MeV}, \qquad c_1^{} = 0.754, \qquad \bar{c}_1 =
0.532.
\end{equation}
Thirdly, in order to determine the empirical values for the $S$-wave kaon
parameters, we use $\Lambda(1405)$, $\Xi(1820)$, and $\Xi(2120)$.%
\footnote{We assume that the spin-parity of $\Xi(2120)$ is $\frac32^+$.
This is based on the observation that its mass is higher than
$\Xi(1820)$ by $300$ MeV, which is the typical energy scale of the
energy difference between the $P$-wave and $S$-wave kaon-soliton systems.
The $\Sigma(1670)\frac32^-$ cannot be used for the fitting process
because it contains the same $\bar{c}_2^{}$ term with $\Lambda(1405)$
and $\Xi(1820)$.}
Then we have
\begin{equation}
\omega_2^{} = 479 \mbox{ MeV}, \qquad c_2^{} = 0.641, \qquad \bar{c}_2 =
0.821.
\end{equation}
This implies that the $S$-wave kaon is very slightly bound with the
binding energy of about 15 MeV.
It also shows that $\bar{c}_1$ $(= 0.532)$ is close to the value of
$c_1^2$ $(=0.569)$, while the difference between $\bar{c}_2$ $(= 0.821)$
and $c_2^2$ $(= 0.411)$ is rather large.
This may indicate that the role of kaon quartic terms would be more
important for the $S$-wave kaons than for the $P$-wave kaons.

The resulting mass spectrum is shown in Table~\ref{tab:mass}.%
\footnote{With the assumption of $\bar{c} = c^2$, we can obtain the
parameters from $\Lambda(1116)$, $\Sigma(1193)$, $\Lambda(1405)$, and
$\Xi(1820)$.
This gives $\omega_1^{} = 223$ MeV, $c_1^{} = 0.604$,
$\omega_2^{} = 492$ MeV, and $c_2^{} = 0.80$.
The resulting mass spectrum is similar to the given in
Table~\ref{tab:mass} within the difference of $40$ MeV at most.}
By considering the simple structure of the mass formula, the resulting
mass spectrum is quite impressive.
In particular, it can explain many of the $\Xi$ resonances reported in
PDG. It also gives a very natural explanation for the low-lying
(one-star resonance) $\Xi(1620)$ and (three-star resonance) $\Xi(1690)$
identifying their spin-parity as $\frac12^-$.

%%%%%%%%%%%%%%%%%% TABLE VI
\begin{table}[t]
\centering
\begin{tabular}{ccc} \hline\hline
Particle Name & Mass (MeV) & Assigned State
\\ \hline
$N$ & $\underline{939}$ \\
$\Delta$ & $\underline{1232}$ \\ \hline
$\Lambda_{1/2^+,0}$ & $\underline{1116}$ & $\Lambda(1116)$ \\
$\Lambda_{1/2^-,1}$ & $\underline{1405}$ & $\Lambda(1405)$ \\
$\Sigma_{1/2^+,0}$ & $1164$ & $\Sigma(1193)$ \\
$\Sigma_{3/2^+,0}$ & $\underline{1385}$ & $\Sigma(1385)$ \\
$\Sigma_{1/2^-,1}$ & $1475$ & $\Sigma(1480)?$ \\
$\Sigma_{3/2^-,1}$ & $1663$ & $\Sigma(1670)$ \\ \hline
$\Xi_{1/2^+,0}$ & $\underline{1318}$ & $\Xi(1318)$ \\
$\Xi_{3/2^+,0}$ & $1539$ & $\Xi(1530)$ \\
$\Xi_{1/2^-,1}$ & $1658(1660)^*$ & $\Xi(1690)?$ \\
$\Xi_{1/2^-,2}$ & $1616(1614)^*$ & $\Xi(1620)?$ \\
$\Xi_{3/2^-,1}$ & $\underline{1820}$ & $\Xi(1820)$ \\
$\Xi_{1/2^+,1}$ & $1932$ & $\Xi(1950)?$ \\
$\Xi_{3/2^+,1}$ & $\underline{2120}$ & $\Xi(2120)?$ \\ \hline
$\Omega_{3/2^+,0}$ & $1694$ & $\Omega(1672)$ \\
$\Omega_{1/2^-,1}$ & $1837$ \\
$\Omega_{3/2^-,1}$ & $1978$ \\
$\Omega_{1/2^+,1}$ & $2140$ \\
$\Omega_{3/2^+,1}$ & $2282$ & $\Omega(2250)?$ \\
$\Omega_{3/2^-,2}$ & $2604$
\\ \hline\hline
\end{tabular}
\caption{Mass spectrum of our model.
The underlined values are used to determine the mass parameters.
The values within the parenthesis are obtained by considering
the mixing effect. The question mark after the particle name means that
the spin-parity quantum numbers are not identified by PDG.}
\label{tab:mass}
\end{table}

From the mass formula, one can derive Lagrangian-independent mass
relations.
We found that the bound-state model does not obey the well-known
Gell-Mann--Okubo mass relation and the decuplet equal spacing rule,
which follow from the quark model including the leading order of the
flavor symmetry breaking.
Instead, it satisfies the modified relations,
\begin{eqnarray}
3\Lambda + \Sigma - 2(N+\Xi) = \Sigma^* - \Delta - (\Omega - \Xi^*),
\nonumber \\
(\Omega - \Xi^*) - (\Xi^* - \Sigma^*) = 
(\Xi^* - \Sigma^*) - (\Sigma^* - \Delta),
\label{eq:mass-rel-1}
\end{eqnarray}
where the symbols denote the masses of the corresponding octet and
decuplet ground states.
These modified mass relations are known to be obtained by including the second
order of the symmetry breaking terms, i.e., $m_s^2$ in terms of
the strange quark mass.
Therefore, these modified relations work better than the original mass
relations. (See, e.g., Ref.~\cite{OW99}.)
In fact, the mass formula of the bound-state model contains the $\bm{J}_K^2$
term.
This term arises not only from the $\bm{\Theta}^2$ term but also from
the $\bm{R} \cdot\bm{\Theta}$ term.
Since $j_K^2$ is nothing but the square of the strangeness,
the mass formula (\ref{eq:mass0}) already takes into account the effect
of the second order of the flavor symmetry breaking terms.
As discussed, e.g. in Ref.~\cite{OW99}, the hyperfine relation holds
even with the $m_s^2$ term, and, therefore, our model satisfies
\begin{equation}
\Sigma^* - \Sigma + \frac32 ( \Sigma-\Lambda) = \Delta - N.
\label{eq:hyp}
\end{equation}

Since the mass relations (\ref{eq:mass-rel-1}) and (\ref{eq:hyp}) are
obtained for the hyperons with the $P$-wave kaons, the same relations should
be true for the hyperons containing the $S$-wave kaons only.
Therefore, those relations are valid by replacing
$\Lambda$, $\Sigma$, $\Sigma^*$, $\Xi$, $\Xi^*$, and $\Omega$ by
$\Lambda_{1/2^-,1}$, $\Sigma_{1/2^-,1}$, $\Sigma_{3/2^-,1}$,
$\Xi_{1/2^+,1}$, $\Xi_{3/2^+,1}$, and $\Omega_{3/2^-,2}$, respectively.
Note that those mass sum rules relate the mixed parity states of
hyperons, i.e., odd-parity $\Lambda$ and $\Sigma$, even-parity $\Xi$,
and odd-parity $\Omega$.

One can derive additional mass sum rules like
\begin{equation}
\Omega_{3/2^+,1} - \Omega_{3/2^-,1} =
\Omega_{1/2^+,1} - \Omega_{1/2^-,1},
\end{equation}
for $\Omega$ resonances.
This shows that the mass differences between the baryons of the same quantum
numbers but of opposite parity, which we call ``parity partners'', are exactly
identical for these baryons.
Although the mass splitting of other parity partner hyperons are not exactly
equal to the above formula, we observe that their mass differences are always
close to $\sim 290$ MeV and the same pattern is clearly seen in the
experimental data.
This is due to the fact that their mass difference starts from the $O(N_c^0)$
term in our mass formula, which is the energy of the bound kaon.
Although $E_{\rm rot}$ contributes to the mass differences of the parity
partners, its effect is $O(N_c^{-1})$ and is about 10\% of the
$O(N_c^0)$ term in our model.
As a result, the equal mass splitting rule between the parity partners
becomes a good approximation.
This is different from the mass splittings between different spin
states of the same parity, which are governed by $E_{\rm rot}$,
the $O(N_c^{-1})$ term.

We now consider the $\bm{R} \cdot (\bm{J}_1^{} - \bm{J}_2^{})$ term in
the mass formula (\ref{eq:mass0}).
As we have discussed before, this term only gives the mixing of the
$\Xi$ resonances, in particular, for $\Xi_{1/2^-,1}$ and $\Xi_{1/2^-,2}$ as
both of them have the same quantum numbers except $j_m^{}$.
By using the state wavefunctions given in Table~\ref{tab:states} we
obtain
\begin{equation}
\langle \Xi_{1/2^-,2} | \bm{R} \cdot \bm{K} | \Xi_{1/2^-,1}
\rangle =
\langle \Xi_{1/2^-,1} | \bm{R} \cdot \bm{K} | \Xi_{1/2^-,2}
\rangle = \frac{\sqrt3}{2}.
\end{equation}
Thus the mixing term in the mass is estimated to be
\begin{equation}
\Delta M = \frac{\sqrt3}{4\mathcal{I}}( c_1^{} - c_2^{}) \approx 9.6
\mbox{ MeV}.
\end{equation}
Therefore, when we define $\Xi_{H,L}$ as
\begin{eqnarray}
\Xi_H &=& \cos\theta\, \Xi_{1/2^-,1} + \sin\theta\, \Xi_{1/2^-,2}, \nonumber \\
\Xi_L &=& -\sin\theta\, \Xi_{1/2^-,1} + \cos\theta\, \Xi_{1/2^-,2},
\end{eqnarray}
we have the masses of $\Xi_H$ and $\Xi_L$ as $1660$ MeV and $1614$ MeV,
respectively, with $\theta \approx 27.5^\circ$.
So the effect of this mixing on the mass spectrum is small and may be
neglected.

\section{Magnetic Moments}

The magnetic moment operator can be written as \cite{OMRS91}
\begin{equation}
\hat\mu = \hat\mu_s + \hat\mu_v,
\end{equation}
where
\begin{eqnarray}
\hat\mu_s &=& \mu_{s,0} R^z + \mu_{s,1} J^z_1 + \mu_{s,2} J^z_2,
\nonumber \\
\hat\mu_v &=& -2 (\mu_{v,0} + \mu_{v,1} n_1 + \mu_{v,2} n_2) D^{33},
\end{eqnarray}
with $D^{33} = -I^z R^z / \bm{I}^2$.
Here, $\mu_{s,0}$ and $\mu_{v,0}$ are the magnetic moment parameters of
the SU(2) sector while $\mu_{s,1}$ and $\mu_{v,1}$ ($\mu_{s,2}$ and
$\mu_{v,2}$) are those for the $P$-wave ($S$-wave) kaon.
The number of $P$-wave and $S$-wave kaons is denoted by $n_1$ and
$n_2$, respectively.
There are totally six parameters and they can be calculated for a given
model Lagrangian \cite{KM89,OMRS91,SSG95,OP96b,NR89,SW03b}.

With the wavefunctions of baryons obtained in the previous Section, it
is straightforward to express the magnetic moments of baryons in terms
of those six parameters.
The results are given in Table~\ref{tab:magmom}.

%%%%%%%%%%%%%%%%%% TABLE VII
\begin{table*}[t]
\centering
\begin{tabular}{cc} \hline\hline
Particle & Magnetic moment \\ \hline
$N$ & $\frac12 \mu_{s0} + I^z \frac43 \mu_{v0}$ \\
$\Delta$ & $\frac32 \mu_{s0} + I^z \frac45 \mu_{v0}$ \\ \hline
$\Lambda_{1/2^+,0}$ & $\frac12 \mu_{s1}$ \\
$\Sigma_{1/2^+,0}$ & $\frac23 \mu_{s0} - \frac16 \mu_{s1} + I^z
\frac23 (\mu_{v0} + \mu_{v1})$
\\
$\Sigma_{3/2^+,0}$ & $\mu_{s0} + \frac12 \mu_{s1} + I^z
(\mu_{v0} + \mu_{v1})$
\\
$\Lambda_{1/2^-,1}$ & $\frac12 \mu_{s2}$
\\
$\Sigma_{1/2^-,1}$ & $\frac23 \mu_{s0} - \frac16 \mu_{s2} + I^z
\frac23 (\mu_{v0} + \mu_{v2})$
\\
$\Sigma_{3/2^-,1}$ & $\mu_{s0} + \frac12 \mu_{s2} + I^z
(\mu_{v0} + \mu_{v2})$
\\ \hline
$\Xi_{1/2^+,0}$ & $ -\frac16 \mu_{s0} + \frac23 \mu_{s1} - I^z
\frac49 (\mu_{v0} + 2 \mu_{v1})$
\\
$\Xi_{3/2^+,0}$ & $ \frac12 \mu_{s0} + \mu_{s1} + I^z
\frac43 (\mu_{v0} + 2 \mu_{v1})$
\\
$\Xi_{1/2^-,1}$ & $ \frac12 \mu_{s0} + I^z
\frac43 (\mu_{v0} + \mu_{v1} + \mu_{v2})$
\\
$\Xi_{1/2^-,2}$ & $ -\frac16 \mu_{s0} + \frac13 (\mu_{s1} +
\mu_{s2}) - I^z \frac49 (\mu_{v0} + \mu_{v1} + \mu_{v2})$
\\
$\Xi_{3/2^-,1}$ & $ \frac12 (\mu_{s0} + \mu_{s1} + \mu_{s2})
+ I^z \frac43 (\mu_{v0} + \mu_{v1} + \mu_{v2})$
\\
$\Xi_{1/2^+,1}$ & $ -\frac16 \mu_{s0} + \frac23 \mu_{s2} - I^z
\frac49 (\mu_{v0} + 2 \mu_{v2})$
\\
$\Xi_{3/2^+,1}$ & $ \frac12 \mu_{s0} + \mu_{s2} + I^z
\frac43 (\mu_{v0} + 2 \mu_{v2})$
\\ \hline
$\Omega_{3/2^+,0}$ & $ \frac32 \mu_{s1}$
\\
$\Omega_{1/2^-,1}$ & $ \frac23 \mu_{s1} - \frac16 \mu_{s2}$
\\
$\Omega_{3/2^-,1}$ & $ \mu_{s1} + \frac12 \mu_{s2}$
\\
$\Omega_{1/2^+,1}$ & $  - \frac16 \mu_{s1} + \frac23 \mu_{s2}$
\\
$\Omega_{3/2^+,1}$ & $ \frac12 \mu_{s1} + \mu_{s2}$
\\
$\Omega_{3/2^-,2}$ & $ \frac32 \mu_{s2}$
\\ \hline\hline
\end{tabular}
\caption{Magnetic moments of baryons. Here, $I^z$ represents the third
component of the isospin.}
\label{tab:magmom}
\end{table*}

The values of the magnetic moment parameters depend on the dynamics of
the soliton-meson system.
Unlike the mass spectrum, we do not have empirical data on the magnetic
moments of odd-parity hyperons.
Therefore, instead of making predictions on the values of baryon
magnetic moments, we suggest several magnetic moment sum rules.

For the ground state baryons, we have
\begin{eqnarray}
\mu(\Sigma^{*+}) - \mu(\Sigma^{*-}) &=& \frac32 \left\{ \mu(\Sigma^+) -
\mu(\Sigma^-) \right\},
\nonumber \\ 
\mu(\Sigma^{+}) + \mu(\Sigma^{-}) &=& \frac43 \left\{ \mu(p) + \mu(n) \right\}
- \frac23 \mu(\Lambda),
\nonumber \\ 
\mu(\Sigma^{*+}) + \mu(\Sigma^{*-}) &=& 2 \left\{ \mu(p) + \mu(n) \right\}
+ 2 \mu(\Lambda),
\nonumber \\
\mu(\Xi^0) + \mu(\Xi^-) &=& - \frac13 \left\{ \mu(p) + \mu(n) \right\} +
\frac83 \mu(\Lambda),
\nonumber \\
\mu(\Xi^{*0}) + \mu(\Xi^{*-}) &=& \mu(p) + \mu(n) + 4 \mu(\Lambda),
\nonumber \\
\mu(\Xi^{*0}) - \mu(\Xi^{*-}) &=& -3 \left\{ \mu(\Xi^{0}) - \mu(\Xi^{-})
\right\},
\nonumber \\
\mu(\Omega) &=& 3 \mu(\Lambda),
\end{eqnarray}
where the symbols represent the octet and decuplet ground state baryons.
The above relations hold by replacing $\Sigma$, $\Sigma^*$, $\Xi$,
$\Xi^*$, $\Omega$ by $\Sigma_{1/2^-,1}$, $\Sigma_{3/2^-,1}$,
$\Xi_{1/2^+,1}$, $\Xi_{3/2^+,1}$, $\Omega_{3/2^-,2}$,
respectively.

%\begin{widetext}

Other interesting sum rules include
\begin{eqnarray}
&&
\mu(\Xi_{3/2^-,1}^0) - \mu(\Lambda_{1/2^-,1})
- \frac12 \left\{ \mu(\Sigma_{1/2^-,1}^+)
- \mu(\Sigma_{1/2^-,1}^-) \right\}
\nonumber \\
&=&
\mu(\Xi_{3/2^+,0}^0) - \mu(\Lambda_{1/2^+,0})
- \frac12 \left\{ \mu(\Sigma_{1/2^+,0}^+)
- \mu(\Sigma_{1/2^+,0}^-) \right\},
\nonumber \\
&&
\mu(\Xi_{3/2^-,1}^-) - \mu(\Lambda_{1/2^-,1}) + \frac12 \left\{
\mu(\Sigma_{1/2^-,1}^+) - \mu(\Sigma_{1/2^-,1}^-) \right\}
\nonumber \\
&=&
\mu(\Xi_{3/2^+,0}^-) - \mu(\Lambda_{1/2^+,0}) + \frac12 \left\{
\mu(\Sigma_{1/2^+,0}^+) - \mu(\Sigma_{1/2^+,0}^-) \right\},
\nonumber \\
&&
\mu(\Xi_{1/2^-,1}^0) + 3 \mu(\Xi_{1/2^-,2}^0) = 
\mu(\Xi_{1/2^-,1}^-) + 3 \mu(\Xi_{1/2^-,2}^-) 
\nonumber \\ &=& 2 \left\{
\mu(\Lambda_{1/2^+,0}) + \mu(\Lambda_{1/2^-,1}) \right\},
\end{eqnarray}

%\end{widetext}

The relations for the excited $\Omega$ baryons are
\begin{eqnarray}
\mu(\Omega_{1/2^-,1}) &=& 
\frac43 \mu(\Lambda_{1116}) - \frac13 \mu(\Lambda_{1405}),
\nonumber \\
\mu(\Omega_{3/2^-,1}) &=& 2 \mu(\Lambda_{1116}) + \mu(\Lambda_{1405}),
\nonumber \\
\mu(\Omega_{1/2^+,1}) &=& -\frac13 \mu(\Lambda_{1116}) + \frac43
\mu(\Lambda_{1405}),
\nonumber \\
\mu(\Omega_{3/2^+,1}) &=& \frac13 \mu(\Lambda_{1116}) + 2
\mu(\Lambda_{1405}),
\end{eqnarray}
by using $\Lambda_{1/2^+,0} = \Lambda(1116)$ and
$\Lambda_{1/2^-,1} = \Lambda(1405)$.

\section{Conclusion}

One of the successes of the bound state approach of the Skyrme model is
that it provides a unified way to explain both $\Lambda(1116)$ and
$\Lambda(1405)$.
The mass difference between the two particles is explained mainly by the
energy difference between the $P$-wave kaon and $S$-wave kaon.
In this paper, we have shown that this feature can be extended to other
low-lying excited states of hyperons.
In particular, in the $\Xi$ baryon spectrum, $\Xi(1820)$ can be explained by
the bound state of soliton and kaons of one in $P$-wave and one in
$S$-wave.
Then the existence of $\Xi(1690)$ and $\Xi(1620)$ baryons of $J^P =
1/2^-$ is required as they are parity-partners of $\Xi(1318)\frac12^+$,
while $\Xi(1690)$ would have $J^P = 1/2^+$ in the
non-relativistic quark model of Chao {\it et al.\/} \cite{CIK81}.

Although the bound-state model has strong resemblance with the
non-relativistic quark model for the ground state baryons, our results show
that the similarities may be lost for excited baryons.
The presence of two low-lying $\Xi$ resonances of $J^P = 1/2^-$, which
we identify as $\Xi(1620)$ and $\Xi(1690)$, comes out naturally in our
model and this is evidently different from other quark model predictions.
In addition, unlike the assumption of Ref.~\cite{AASW03}, we do not need
to require the existence of very low mass nucleon and $\Lambda$
resonances.
We also note that the spin-parity quantum numbers of $\Xi(1620)$ and
$\Xi(1690)$ of our model are consistent with Ref.~\cite{ROB02} and
with Ref.~\cite{GLN04}.
We also found that the bound-state model predicts $\Xi$ resonance with
$J^P = 1/2^+$ at a mass of around 1930 MeV.
This would be a good candidate for the observed three-star resonance
$\Xi(1950)$.
Therefore, it is quite different from the quark model prediction, e.g., of
Ref.~\cite{FP72}, where $\Xi(1950)$ was interpreted to have $J^P = 5/2^-$
as the Gell-Mann--Okubo mass relation for $J^P = 5/2^-$ states
[with $N(1675)$, $\Sigma(1765)$, and $\Lambda(1830)$]
requires a $\Xi$ at a mass of around 1960 MeV.
(See Ref.~\cite{GP05b} for a recent discussion.)
It would be interesting to note that the Skyrme model study for radially
excited states of Ref.~\cite{Weigel04} suggests $J^P=\frac12^+$ for the
$\Xi(1950)$, although $\Xi(1690)$ was speculated to be a $J^P=\frac32^-$
state.

The quark-based models \cite{CIK81,GR96b,BIL00} predict that the lowest
$\Omega(\frac12^-)$ is degenerate in mass with the lowest $\Omega(\frac32^-)$.
However, this degeneracy is no longer valid in the bound state model which
predicts that the mass difference between the two $\Omega$ resonances is
around $140$ MeV.
But our model predicts that the first excited $\Omega$ resonance would
have $J^P = 1/2^-$ as in the quark models.

In $\Sigma$ hyperon resonances, we found that $\Sigma(1670)\frac32^-$
fits well as the parity partner of $\Sigma(1385)$.
And our result for $\Sigma(\frac12^-)$ strongly implies that the
one-star-rated $\Sigma(1480)$, which was recently confirmed by COSY
experiment \cite{COSY-05b}, would have $J^P = \frac12^-$ being the
parity partner of $\Sigma(1193)$.
This should be compared with other model predictions on $\Sigma(1480)$
\cite{HS78,AASW03,OKL03b,GLN04}.

Of course, as mentioned before, the bound-state model cannot be applied to
all hyperon resonances.
In particular, in the Skyrme model, $Y\pi$ resonances should be explored
by introducing fluctuating pion fields explicitly \cite{SWHH89} and the
resonances above the threshold should be treated in a different manner
\cite{PRM04,Scoccola90}.
However, the pattern observed in the empirical mass spectrum of
low-lying hyperon resonances, i.e., (approximately) equal spacings between
parity partners, supports the bound-state approach of the Skyrme model,
and distinguishes this model from the other phenomenological models.

Therefore, verifying the spin-parity quantum numbers of $\Xi(1690)$ and
$\Xi(1950)$ is important to understand the structure of excited hyperons.
In addition, confirming the existence of the one-star $\Xi(1620)$
resonance \cite{BGKS77,HACN81}
is also highly desirable as well as searching for low-lying
$\Omega$ baryons.
Theoretically, more investigation on the role of the $K^*$ vector
mesons in the dynamics of the meson-soliton systems and on the
anharmonic corrections to $\Xi$ and $\Omega$ hyperon resonances is required.
In particular, the hyperon resonances in the pion-hyperon channel should
be examined for more realistic description of the hyperon spectrum.

\acknowledgments

It is a great pleasure to acknowledge fruitful discussions with
B.-Y. Park who pointed out the equal mass spacings in hyperon spectrum.
Useful discussions with and encouragements from C.~M. Ko, T.-S.~H. Lee,
D.~P. Min, and K.~Nakayama are also gratefully acknowledged.
This work was supported in part by the COSY Grant No. 41445282
(COSY-58), the US National Science Foundation under Grant No.
PHY-0457265, and the Welch Foundation under Grant No. A-1358.

%\bibliographystyle{/home/yoh/tex/macros/bibtex/h-physrev4}
%\bibliography{/home/yoh/tex/refer/biba-j,/home/yoh/tex/refer/bibk-z}
%\end{document}

\end{document}